\documentclass[conference]{IEEEtran}
\IEEEoverridecommandlockouts
\usepackage{cite}
\usepackage{amsmath,amssymb,amsfonts}
\usepackage{algorithmic}
\usepackage{graphicx}
\usepackage{textcomp}
\usepackage{xcolor}
\usepackage{url}
\usepackage{array}
\usepackage{enumitem}
\def\BibTeX{{\rm B\kern-.05em{\sc i\kern-.025em b}\kern-.08em
    T\kern-.1667em\lower.7ex\hbox{E}\kern-.125emX}}
\begin{document}

\title{LLM-Based Approach for Enhancing Maintainability of Automotive Architectures\\

\thanks{
This work has received funding from the European Chips Joint Undertaking under Framework Partnership Agreement No 101139789 (HAL4SDV) including the national funding from the Federal Ministry of Research, Technology and Space of Germany under grant number 16MEE00471K. The responsibility for the content of this publication lies with the authors.
}
}

\author{
\IEEEauthorblockN{Nenad Petrovic}
\IEEEauthorblockA{
\textit{Chair of Robotics, Artificial} \\ 
\textit{Intelligence and Real-Time Systems} \\
\textit{Technical University of Munich}\\
Munich, Germany \\
nenad.petrovic@tum.de}
\and
\IEEEauthorblockN{Lukasz Mazur}
\IEEEauthorblockA{
\textit{Chair of Robotics, Artificial} \\ 
\textit{Intelligence and Real-Time Systems} \\
\textit{Technical University of Munich}\\
Munich, Germany \\
lukasz.mazur@tum.de}
\and
\IEEEauthorblockN{Alois Knoll}
\IEEEauthorblockA{
\textit{Chair of Robotics, Artificial} \\ 
\textit{Intelligence and Real-Time Systems} \\
\textit{Technical University of Munich}\\
Munich, Germany \\
knoll@in.tum.de}
}

\maketitle

\begin{abstract}
There are many bottlenecks that decrease the flexibility of automotive systems, making their long-term maintenance, as well as updates and extensions in later lifecycle phases increasingly difficult, mainly due to long re-engineering, standardization, and compliance procedures, as well as heterogeneity and numerosity of devices and underlying software components involved. In this paper, we explore the potential of Large Language Models (LLMs) when it comes to the automation of tasks and processes that aim to increase the flexibility of automotive systems. Three case studies towards achieving this goal are considered as outcomes of early-stage research: 1) updates, hardware abstraction, and compliance, 2) interface compatibility checking, and 3) architecture modification suggestions. For proof-of-concept implementation, we rely on OpenAI's GPT-4o model. 
\end{abstract}


\section{Introduction}

When extending an existing software system to meet new requirements, architects and developers often face a critical decision: whether to leverage pre-existing software components (e.g., open-source libraries) to partially fulfill the new requirements, necessitating integration efforts to ensure compatibility, or to implement the required functionality from scratch. To make an informed design decision, a comprehensive analysis is essential to evaluate and compare the efforts and risks associated with both approaches. This task is inherently complex and time-consuming, requiring a deep understanding of the system and the analysis of multiple specification documents and source code.

On the other side, changes involving adding, removing, or modifying the existing hardware setup of various systems also require additional actions and efforts. In that context, various additional actions have to be considered, such as driver installation, middleware-related and configuration file modifications, as well as making sure that both system and application software work properly after the change. However, such activities might require specific expertise with significant cognitive load involved, as well as effort and time in case when performed manually. Therefore, hardware abstraction and its integration within the system represent one more element that has a huge impact on system flexibility.

When it comes specifically to the automotive industry, long re-engineering cycles and comprehensive standardization even further affect the flexibility of underlying products~\cite{a0}\cite{a1}. Apart from purely technical challenges (such as data format compatibility between components), compliance in automotive additionally makes the extension and/or modification of the system's hardware or software configuration more complex and time-consuming \cite{a0}. However, one of the possible approaches to increase the competitiveness of the European automotive industry is the reduction of time and effort needed for product evolution~\cite{a2}, making the products easily maintainable and customizable from a long-term perspective~\cite{a1}.

The primary objective of this work is to explore the potential of Large Language Models (LLMs) in supporting engineers with tasks related to system evolution, with a focus on the automotive domain. It is known that LLMs exhibit strong potential in use cases relying on text summarization, analysis, and generation, which could possibly reduce the cognitive load, effort, and time needed for supporting activities that would make automotive systems more flexible. In this context, several different aspects will be taken into account. In the first step, it will be examined how the change of hardware component would affect the overall system from the perspective of the reference architecture, enabling or disabling certain specific usage scenarios. For this purpose, we leverage the synergy of LLMs with model-driven engineering (MDE). Afterwards, the focus will be on the more narrow task of analyzing interface compatibility between the new software component and the existing system. Subsequently, the scope is broadened to investigate how LLMs can suggest various architectural modifications required to meet new requirements with respect to standardization-related constraints, both with and without the proposed new component. The generated analyses are intended to accelerate the integration of new software components and facilitate more informed decision-making regarding the system's evolution to meet new requirements, leveraging existing software elements.

\section{Background and Related Works}

While the existing visions of LLM-empowered operating systems targeting the automotive domain also cover aspects of hardware management and automated software updateability~\cite{b1}, to the best of our knowledge, there is still no reference implementation or publicly available example of such a solution. A promising proof-of-concept implementation of a flexible, customizable OS aiming IoT use cases was shown in~\cite{b2}. However, in some of our previous works, we have covered several aspects towards making LLM-driven automotive system evolution possible. In~\cite{b3} and~\cite{b4}, we introduced an approach leveraging the synergy of LLMs with formally grounded methods based on MDE as an intermediary step before code and test generation starting from textual requirements and scenario descriptions. In~\cite{b3}, we also propose an automotive system metamodel and interface specification as a foundation for hardware abstraction. Additionally, relying on model-based representation, it is possible to enable verifiability and compliance-related checks of created automotive system models with respect to Object Constraint Language (OCL) rules constructed using LLM starting from reference architecture and standardization documents~\cite{b5}. Moreover, in~\cite{b6} we also take into account visual sources of requirements specification and system representation, leveraging Vision Language Models (VLM) for the purpose of identifying system updates that might result from changes in diagrams describing the automotive system. However, in this paper, we go a step further by focusing on system consistency while performing modifications and updates, taking into account the aspects of hardware abstraction and automating the underlying operations relying on LLMs.

\section{Case Studies}
\subsection{Case Study 1: System Updates, Hardware Abstraction and Compliance}
Illustration depicting the proposed workflow that aims to provide convenient automotive system updates leveraging LLM-driven hardware abstraction and compliance checking is given in Fig. 1.

\begin{figure*}[htbp]
\centering
\includegraphics[width=0.8\textwidth]{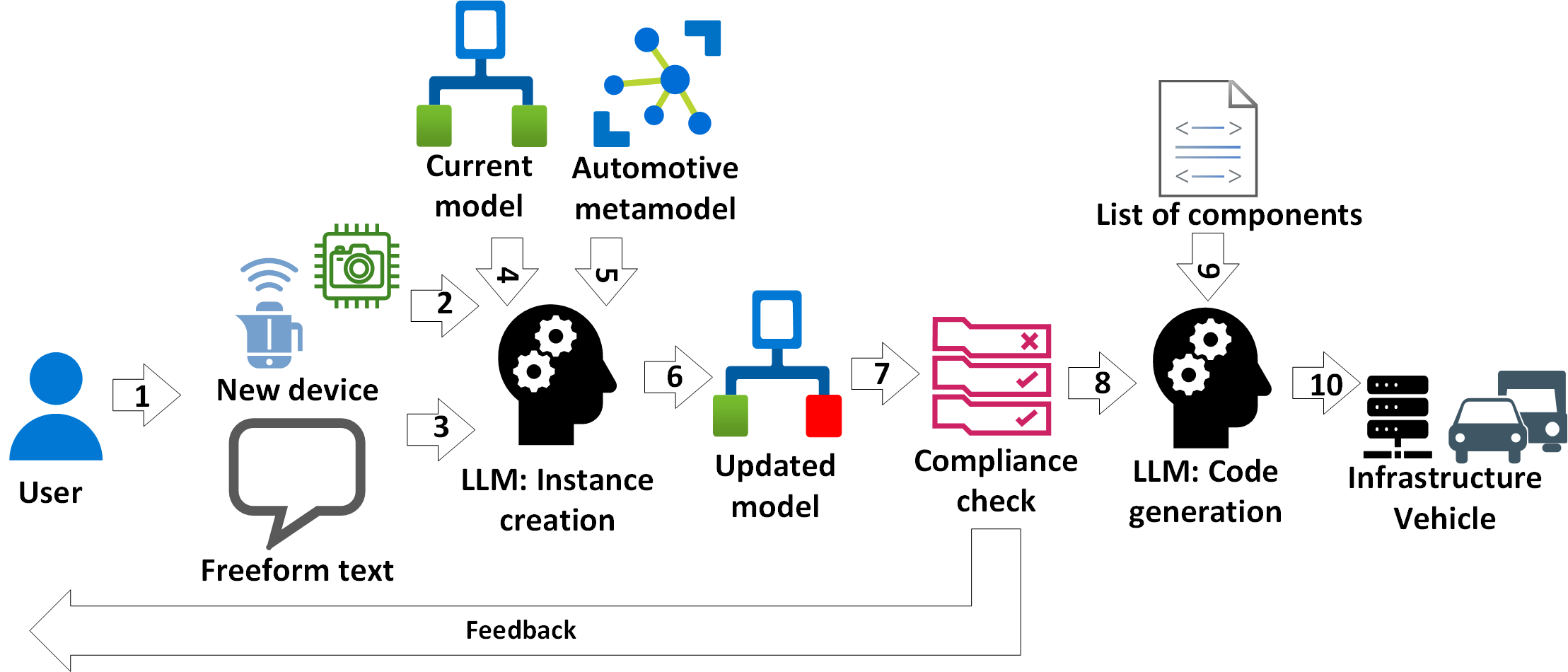}
\caption{Updates and compliance workflow: 1) user input; 2) new device description; 3) usage scenario for new device; 4) XMI model instance; 5) Ecore metamodel; 6) updated XMI model instance; 7) pre-defined set of OCL-rules; 8) check pass; 9) drivers, ROS topics, Docker containers; 10) target commands.}
\label{c1}
\end{figure*}

First, we assume that the user wants to add a new device to the existing vehicular system, such as a sensor, actuator, or application-specific hardware accelerator. Moreover, it is also considered that the specification of the new device is provided (either entered by the user or retrieved from the device descriptor automatically), together with freeform text that aims to describe the intended purpose of the new device, which will be added. In this step, we take into account the current system state expressed as a model instance, as well as the underlying metamodel (like the one from~\cite{b3}) in order to construct the updated system model based on user-provided text and device description. For this purpose, we rely on our previous work on LLM-based instance model creation from~\cite{b7}. As an outcome, the model-based representation of the updated ("to-be") system is constructed. The implementation of metamodel-related aspects is based on Eclipse Modeling Framework (EMF)~\cite{b8} and its Ecore framework~\cite{b9}, while specific model instances are represented in XMI format.
On the other side, a set of design space constraints is taken into account to be checked against the updated system's model-based representation. For that purpose, OCL rules~\cite{b10} based on assumed reference architecture (as described in~\cite{b3}) are taken into account. In our case, the goal of such rules is to identify whether the updated system configuration is capable of performing the desired user scenario. If all the constraints hold within the updated instance model representing the "to-be" system, then we are able to continue with LLM-based code generation. Code generation process, apart from updated system model, also takes into account catalog of pre-defined components which are assumed to be available: 1) list of device-specific drivers for target host architecture 2) publish/subscribe message topics for ROS2 middleware 3) Docker containers list. The commands generated as an outcome are further executed against the target infrastructure~\cite{b11}.

\subsection{Case Study 2: Interface Compatibility Checking}

The proposed approach utilizes LLMs as a Software Interface Compatibility Checker, which processes the specifications of the existing system, new requirements, and new component specifications. The specifications of both the existing system and the new component can comprise various artifacts such as documentation, software models, and source code, each providing complementary perspectives on the system. For instance, documentation might reveal dependencies, such as third-party libraries, that are not explicitly mentioned in the source code. The new requirements may involve modifications to existing software requirements or the specification of new functionalities achievable with the new component.

Once the inputs are processed, the Interface Compatibility Checker generates a thorough analysis of compatibility across all relevant interface aspects. This includes matching data types, semantic alignment of interface data and behavior, analysis of pre- and post-conditions, verification of interface invariants, and the communication mechanisms in use. The resulting analysis can then be used to evaluate the degree of incompatibility between the new component and the existing system, and to decide if source code modifications are necessary and feasible.

\begin{figure*}[htbp]
\centering
\includegraphics[width=\textwidth]{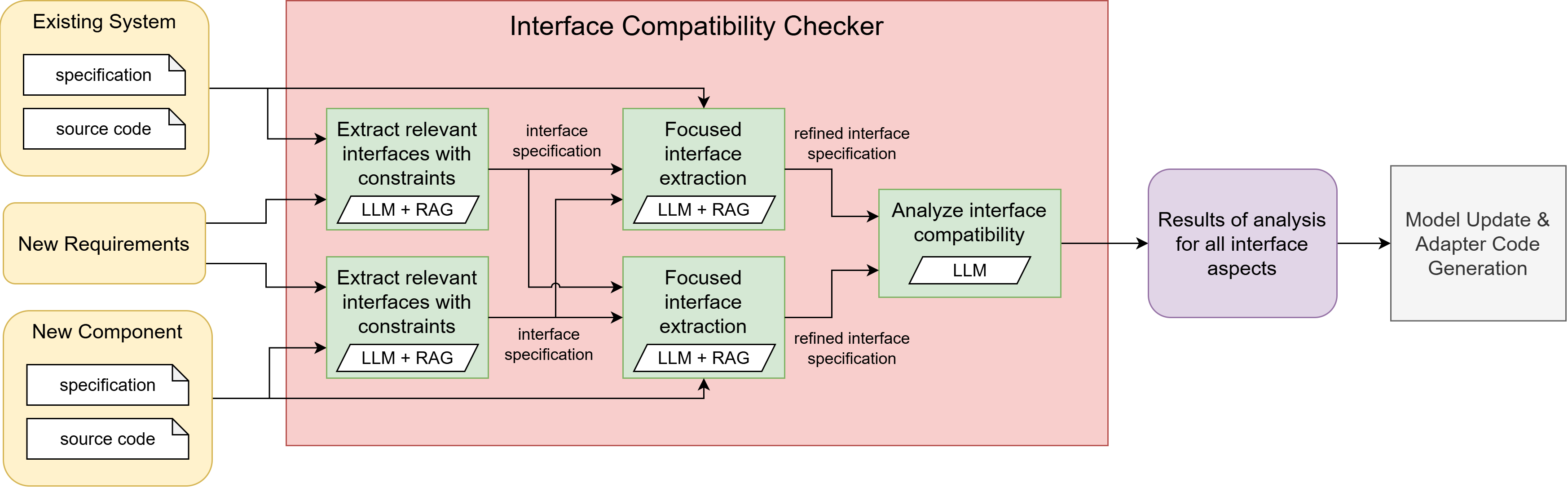}
\caption{Workflow of the Software Interface Compatibility Checker.}
\label{workflow_interface_compatibility_checker}
\end{figure*}

The workflow of the Interface Compatibility Checker is illustrated in Fig.~\ref{workflow_interface_compatibility_checker}.

\subsection{Case Study 3: Architecture Modification Suggestions}

To address the broader tasks of supporting software architecture evolution, we propose a Software Architect Assistant. This assistant processes the same inputs as the Interface Compatibility Checker (specifications of the existing system, new requirements, and new component specifications), but additionally incorporates input from the software architect. These inputs can include design preferences or additional knowledge not present in the specifications, such as planned transitions to new design patterns (e.g., using private databases for microservices) or insights about future functionalities planned in the roadmap.

As a result, the Architect Assistant generates a list of analyzed architecture modification options, highlighting the benefits and drawbacks of each. This enables the software architect to select the most suitable option. The assistant also implicitly analyzes interface compatibility between the new software component and the existing system, provides various options for integrating the new component, and compares these with options for addressing the requirements without the new component. After selecting the preferred architecture modification option, a more detailed plan can be prepared (e.g., feature or user story definition) to serve as a basis for implementation. While these steps (selection, detailed planning, and implementation) can also be automated using LLMs, they are beyond the scope of the current work.

\begin{figure*}[htbp]
\centering
\includegraphics[width=\textwidth]{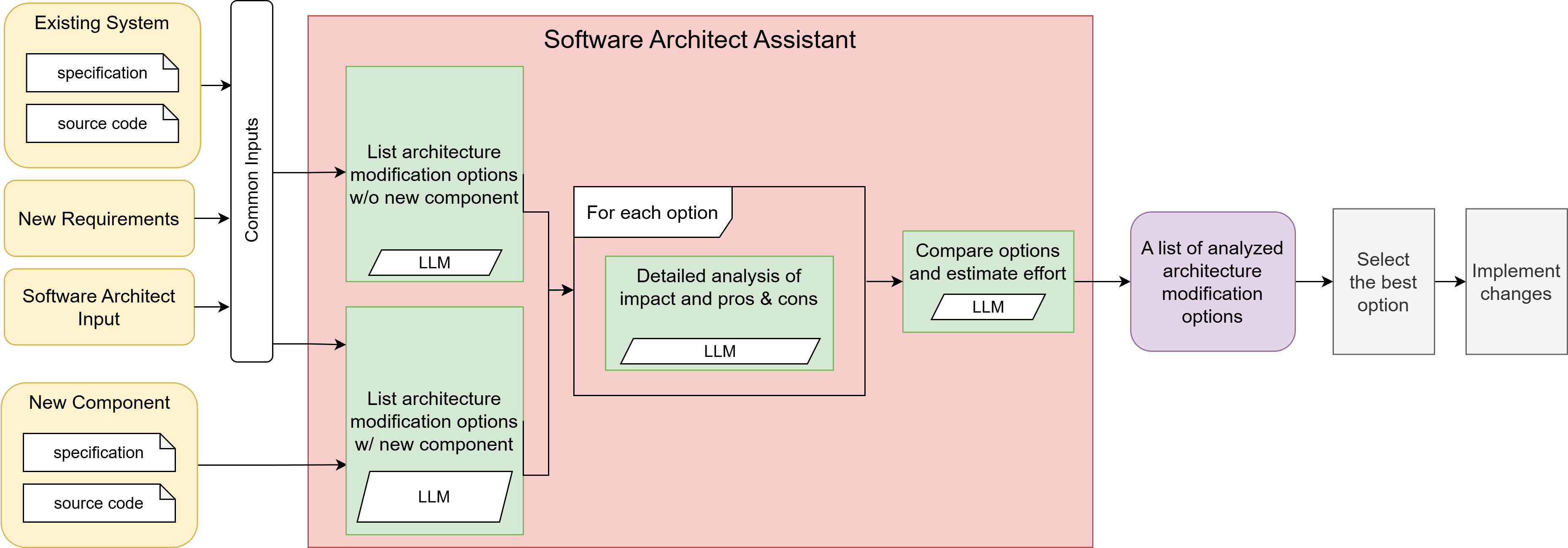}
\caption{Workflow of the Software Architect Assistant.}
\label{workflow_architect_assistant}
\end{figure*}

The workflow of the Architect Assistant is illustrated in  Fig.~\ref{workflow_architect_assistant}.

\section{Experiment Scenarios}
\subsection{Hardware Abstraction and Compliance}
Scenarios for this case study are inspired by reference architecture examples for autonomous and assisted driving capabilities from~\cite{b12}. A summary of these scenarios is given in Tab.~\ref{abstraction_compliance_tab}.

First, let us assume that we target a vehicle configuration with two cameras (2x 8.3MP), five radars, and ten ultrasonic sensors, plus potentially one driver monitoring camera (1x 8.3MP). This specification is used as the basis for OCL rule generation. In this experiment, we want to implement supervised parking with such a configuration based on the reference architecture from~\cite{b12}.
However, let us assume that our current system lacks one camera. If we add a camera with 2.1MP resolution, then the OCL rule check will fail and code generation won't happen. Otherwise, if the added camera has resolution equal to or higher than 8.3MP, we will proceed to command generation, involving installation of driver script, subscription to ROS2 topics, and running the Docker container for parking assistance.

\begin{table}[htbp]
\caption{Updates and Hardware Abstraction Examples}
\begin{center}
\begin{tabular}{|p{0.11\textwidth}|p{0.11\textwidth}|p{0.06\textwidth}|p{0.12\textwidth}|}
\hline
\textbf{Target} & \textbf{Current}& \textbf{New \mbox{device}}& \textbf{Expected \mbox{Outcome}} \\
\hline
\mbox{2x 8.3MP camera,} \mbox{5x radar,} \mbox{10x ultrasonic,} \mbox{1x 8.3MP driver} \mbox{\ camera [optional]} &
\mbox{1x 8.3MP camera,} \mbox{5x radar,} \mbox{10x ultrasonic}  &
Camera model c0 - 2.1MP  & 
Camera \mbox{resolution} \mbox{too low} \\
\hline
\mbox{2x 8.3MP camera,} \mbox{5x radar,} \mbox{10x ultrasonic,} \mbox{1x 8.3MP driver} \mbox{\ camera [optional]} &
\mbox{1x 8.3MP camera,} \mbox{5x radar,} \mbox{10x ultrasonic}  &
Camera model c1 - 8.3MP  & 
\mbox{Install c1,} \mbox{run docker parking,} \mbox{publish topic} \mbox{camera2parking} \\
\hline
\end{tabular}
\label{abstraction_compliance_tab}
\end{center}
\end{table}

\subsection{Interface Compatibility Inspired by Autoware HAL}

Autoware is an open-source autonomous driving software stack designed to operate across a variety of vehicle platforms. Its modular architecture includes a Hardware Abstraction Layer (HAL), which bridges high-level autonomous control and vehicle-specific hardware and protocols. A critical component of this HAL is the Vehicle Interface~\cite{b13}, which translates motion commands and status reports between Autoware’s internal message formats and the target vehicle’s control systems.

Due to the diversity of vehicle brands and communication standards, interfacing often requires custom adapters. One representative example is the PACMod3 ROS driver~\cite{b14}, which communicates with vehicles using the PACMod drive-by-wire system via CAN protocols. The driver standardizes actuation commands across supported platforms and exposes them through ROS 2 messages such as \texttt{pacmod3\_msgs/SteeringCmd}. To integrate with Autoware, additional adapters - like \texttt{pacmod\_interface} - translate these messages to Autoware-specific types such as \texttt{autoware\_control\_msgs/Control}.

Given this adapter-based integration, it is plausible that several types of interface incompatibilities had to be resolved during the development of \texttt{pacmod\_interface}, including:

\begin{itemize}

\item \textit{Data type mismatch}: 
For example, steering command field \texttt{SteeringCmd.rotation\_rate} is defined as a \texttt{float64}, while the corresponding \texttt{Lateral.steering\_tire\_rotation\_rate} uses \texttt{float32}. Although compatible in some contexts, such mismatches can lead to precision loss or runtime errors if not handled explicitly.

\item \textit{Semantic mismatch}:
The two interfaces may define steering direction conventions differently (e.g., clockwise vs. counterclockwise), leading to unintended behavior unless properly transformed.

\item \textit{Communication mechanism mismatch}:
While both systems typically rely on publish/subscribe patterns, mismatches can occur in expected message frequency or when integrating systems using different communication paradigms, such as service/client.

\end{itemize}

To support the resolution of such incompatibilities, an LLM-based approach, such as Software Interface Compatibility Checker, could be employed to analyze interfaces, identify potential issues, and suggest or generate adapter code to facilitate correct integration.

Based on these incompatibility types, a set of simplified ROS 2 evaluation scenarios was constructed. These setups abstract away from real-world vehicle hardware and instead focus on minimal message exchange configurations designed to isolate specific incompatibility types. Each scenario includes a data-producing node (publisher or service server) and a~data-consuming node (subscriber or client), as summarized in Table~\ref{interface_compatibility_scenarios_tab}.

\begin{table}[htbp]
\caption{Interface Compatibility Scenarios}
\begin{center}
\begin{tabular}{|>{\raggedright\arraybackslash}p{0.1\textwidth}
                |>{\raggedright\arraybackslash}p{0.15\textwidth}
                |>{\raggedright\arraybackslash}p{0.15\textwidth}|}
\hline
\textbf{Scenario} & \textbf{Setup} & \textbf{Expected Outcome} \\
\hline
Data type mismatch &
Published rotation rate as \texttt{float32}; subscriber expects \texttt{float64}. &
Precision loss identified; explicit conversion suggested. \\
\hline
Semantic mismatch &
Both publisher and subscriber use \texttt{float32} for rotation rate; documentation describes different steering conventions. &
Mismatch identified; transformation between conventions recommended. \\
\hline
Communication mechanism mismatch &
Rotation rate command published at 50~Hz; subscriber expects 100~Hz. &
Rate mismatch identified; potential solutions include interpolation or resampling. \\
\hline
\end{tabular}
\label{interface_compatibility_scenarios_tab}
\end{center}
\end{table}

\subsection{Architect Assistant: Simple Face Detection System}

The Architect Assistant approach was evaluated using a~simple face detection system for a webcam built with ROS 2. The setup consists of three nodes: a ROS 2 driver for USB cameras (\texttt{usb\_cam}~\cite{b15}), a node for pre-processing images (\texttt{image\_filter}), and a node for performing face detection (\texttt{face\_detector}). The \texttt{image\_filter} node subscribes to raw images published by \texttt{usb\_cam}, crops them to the desired size, and then publishes them. The \texttt{face\_detector} node subscribes to these cropped images, performs face detection using the OpenCV face detector based on the Haar Cascade classifier~\cite{b16}, and publishes an image with detected faces for viewing in the client application (e.g., \texttt{rviz2}).

In the experiment, the existing system used \texttt{image\_filter\_small} to crop the image to a size of 640x480. The new requirement was to support a~resolution of 1280x720, and a modified image filter node (\texttt{image\_filter\_large}) was provided as a new component to perform cropping to this size. The goal was to determine if the Architect Assistant could correctly suggest different options for modifying the existing implementation of \texttt{image\_filter\_small}, as well as suggest replacing \texttt{image\_filter\_small} with \texttt{image\_filter\_large} after checking interface compatibility.

The specifications for all components in the existing system and the new component were generated in Markdown format and reviewed before being used as input to the Architect Assistant. The source code of the existing system and the new component were provided as single strings consisting of the folder structure for each component, followed by the concatenated contents of each file. The software architect's input consisted of general guidelines for the Assistant, such as adhering to the single responsibility principle and the open/closed principle.

\section{Results overview}
In this section, we present the summary of the achieved outcomes for the previously presented case studies, as shown in Table \ref{tab:rel4}. For each of them, we show the crucial steps, the main instructional content embedded in the prompts, as well as performance in terms of execution time and percentage of correctness, given as average based on 10 executions. We used OpenAI's GPT-4o LLM in all of the presented experiments. Based on the achieved values, it can be noticed that execution time grows with the size of textual inputs considered, causing larger token consumption in such cases, as expected. Additionally, it can be noticed that adoption of synergy with MDE with formal representations as intermediate steps has two main benefits: 1) usage of compact model instance representations instead of full source code resulting with reduced token consumption, 2) higher percentage of correct answers due to verifiable model instances and formal rules adopted.

\begin{table*}[htbp]
\caption{Experiment results summary}
\begin{center}
\scriptsize
\begin{tabular}{|c
                |>{\raggedright\arraybackslash}p{7.6cm}
                |>{\raggedright\arraybackslash}p{3.3cm}
                |>{\raggedright\arraybackslash}p{2.0cm}
                |>{\raggedright\arraybackslash}p{2.0cm}|}
\hline
\textbf{Case} & \textbf{Prompts/steps} & \textbf{Inputs} & \textbf{Execution time [s]} & \textbf{Correctness} \\
\hline
1  & 
Step 1: OCL rule generation
\begin{itemize}[leftmargin=*]
    \item Generate OCL rules based on reference requirements.
\end{itemize}

Step 2: Update model
\begin{itemize}[leftmargin=*]
    \item Update current system model instance in order to add new device satisfying given new requirements, taking into account metamodel.
\end{itemize}
Step 3: Validate model

Step 4: Generate code
\begin{itemize}[leftmargin=*]
    \item Generate driver installation commands for new device.
    \item Generate docker container run command based on template.
    \item Subscribe to ROS 2 topics based on list.
\end{itemize}

&  
\begin{itemize}[leftmargin=*]
    \item Current model
    \item Metamodel
    \item New device specification
    \item New requirements
    \item Reference requirements
    \item Docker command template 
    \item List of ROS 2 topics
\end{itemize}

 & 38.55 &  
 9/10 (Code generation) 
 
 10/10 (Decision) \\
\hline
2
& 
Step 1: Extraction
\begin{itemize}[leftmargin=*]
    \item Extract all parts of the system’s interfaces relevant to the new requirement.
    \item Extract all parts of the component’s interfaces relevant to the new requirement.
\end{itemize}

Step 2: Refinement
\begin{itemize}[leftmargin=*]
    \item Refine the system’s interface specification to ensure it addresses all aspects required for integrating the new component.
    \item Refine the component’s interface specification to ensure it supports seamless integration into the existing system.
\end{itemize}

Step 3: Compatibility Analysis
\begin{itemize}[leftmargin=*]
    \item Analyze all aspects of both interfaces to assess their compatibility.
\end{itemize}
& 
\begin{itemize}[leftmargin=*]
    \item Existing system specification
    \item Existing system source code
    \item New component specification
    \item New component source code
    \item New requirement
\end{itemize}
&
Mismatch cases:

89.08 (Data type)

90.05 (Semantic)

101.64 (Comm. mechanism)
& 
Mismatch cases:

9/10 (Data type)

3/10 (Semantic)

9/10 (Comm. mechanism)
\\
\hline
3
&
Step 1: Architectural Options Identification
\begin{itemize}[leftmargin=*]
    \item Describe options for modifying the existing system’s architecture to fulfill the new requirement without introducing a new software component.
    \item Describe options for modifying the architecture by introducing a new software component to fulfill the requirement.
\end{itemize}

Step 2: Detailed Analysis of Modifications
\begin{itemize}[leftmargin=*]
    \item For each proposed modification, perform an analysis including a list of required changes and advantages and disadvantages of the approach.
\end{itemize}

Step 3: Effort and Impact Estimation
\begin{itemize}[leftmargin=*]
    \item Compare the proposed modifications and provide relative estimates of the implementation effort and impact for each option.
\end{itemize}
& 
\begin{itemize}[leftmargin=*]
    \item Existing system specification
    \item Existing system source code
    \item New component specification
    \item New component source code
    \item New requirement
    \item Software architect input
\end{itemize}
& 116.41 &  10/10 \\
\hline
\end{tabular}
\label{tab:rel4}
\end{center}
\end{table*}

\section{Conclusion}
Based on the initial outcomes, it can be concluded that LLMs exhibit strong potential when it comes to activities related to the extendibility of automotive systems, which is beneficial for increased flexibility and opens many possibilities for product customizations and updates even in later phases of the product lifecycle. Regarding future work, our focus would be to evaluate various LLM solutions. Additionally, the synergy of agentic AI approaches complementing LLMs will be considered, taking into account vehicle maintenance scenarios as well. Finally, we will also consider the adoption of LLMs in order to provide support for the generation of Vehicle Signal Specification (VSS)~\cite{b17} based data model starting from either textual specification or diagrams, as well as the underlying mechanisms for validation of changes which might result from hardware and software updates.


\vspace{12pt}

\end{document}